# Ultrafast magneto-photocurrents in GaAs: Separation of surface and bulk contributions

Christian B. Schmidt,[1,a)] Shekhar Priyadarshi,[1] Sergey A. Tarasenko,[2] and Mark Bieler[1]

[1] Physikalisch-Technische Bundesanstalt, Bundesallee 100, 38116 Braunschweig, Germany

[2] Ioffe Institute, Politechnicheskaya 26, 194021 St. Petersburg, Russia

We induce ultrafast magneto-photocurrents in a GaAs crystal employing interband excitation with femtosecond laser pulses at room temperature and non-invasively separate surface and bulk contributions to the overall current response. The separation between the different symmetry contributions is achieved by measuring the simultaneously emitted terahertz radiation for different sample orientations. Excitation intensity and photon energy dependences of the magneto-photocurrents for linearly and circularly polarized excitations reveal an involvement of different microscopic origins, one of which we believe is the inverse Spin-Hall effect. Our experiments are important for a better understanding of the complex momentum-space carrier dynamics in magnetic fields.

[a)] Author to whom correspondence should be addressed. Electronic mail: christian.b.schmidt@ptb.de

All-optical excitation of semiconductors might produce macroscopic currents that depend on the state of optical polarization.[1–4] These currents are typically linked to the photogalvanic effect and, thus, require a reduced point group symmetry, which can be provided by the crystal itself,[1] deformation or nanostructuring[2] of the crystal, and the surface.[5,6] The point group symmetries are further reduced in external magnetic fields leading to additional currents.[6–11] Besides the well known Hall effect, which simply redirects already existing currents, also non-Hall currents occur.[7–13] In the following we refer to this current type as magneto-photocurrent (MC). MCs may also originate from symmetry breaking due to the surface,[7,9] the crystal structure,[8] or nanostructuring of the crystal.[13]

Previous MC studies in bulk GaAs demonstrated that MCs induced by linearly polarized light are linked to asymmetric relaxation of an anisotropic carrier distribution.[7,8] In contrast, MCs generated by circularly polarized light are related to spin splitting[14] and can even be linked to the inverse spin-Hall effect.[9,12,15,16] Thus, MCs provide a possibility to study very different physical phenomena and their dynamics in semiconductors. However, in order to do so, two experimental challenges have to be addressed: (i) It is necessary to separate surface and bulk contributions to the MCs.[8,9] (ii) In order to gain access to the ultrafast current dynamics a detection technique providing sub-picosecond time resolution has to be employed instead of previously used time-integrated detection techniques.[7–9]

To address the aforementioned problems we used ultrafast laser pulses to optically excite a bulk GaAs crystal ($E_g = 1.42$ eV) to induce the MCs at room temperature and measured the simultaneously emitted terahertz (THz) radiation employing electro-optical sampling.[17,18] This way we are able to distinguish between surface and bulk contributions to the MCs in a non-invasive way by taking advantage of their different symmetry properties. We observe surface MCs (SMCs) induced by linearly and circularly polarized light. However, bulk MCs (BMCs) are only observed for circularly but not for linearly polarized excitation. A linear behavior of all MCs on magnetic fields up to ±1.2T is obtained. While the MC dependence on optical intensity shows diverse saturation effects, the dependence on optical photon energy is non-monotonic and complex. These observations clearly show different microscopic contributions to the MCs and cannot be explained by existing time independent theories.[10,12,14]

*Experimental and theoretical basics* - A detailed outline of our pump-probe setup is shown in Fig. 1(a). As light source a 76 MHz Ti:Sa laser with a temporal and spectral pulse width of 150 fs and 10 meV, respectively, was used. The pump beam was focused on a 450μm thick (001)-orientated GaAs wafer with a spot diameter of 220 μm, corresponding to peak intensities up to 120 MW cm$^{-2}$. Two off-axis parabolic mirrors were employed to collect the THz radiation emitted from the backside of the sample and to focus it on a 0.5 mm thick ZnTe crystal used for electro-optic sampling. With a THz polarizer and proper orientation of the ZnTe crystal in the setup we ensured exclusive detection of THz radiation from currents flowing (anti-) parallel to the magnetic field. A low pass filter (LPF) between the two off axis parabolic mirrors ensured, that no residual pump beam reached the detection in case of below band gap excitation. The magnetic field was obtained by an electro-magnet with field strengths ranging between ±1.2 T at the sample position.

In the following we give a phenomenological description of MCs and point out how surface and bulk contributions can be separated. Theoretically, MCs are described as a second-order response to the electric field of the optical beam. Assuming a linear dependence on the static magnetic field, one obtains for MCs induced by ultrafast laser pulses

$$j_{MC_\alpha}(t) = \int_{-\infty}^{t} dt' \int_{-\infty}^{t'} dt'' \sum_{\beta\gamma\delta} \Phi_{\alpha\beta\gamma\delta}(t-t', t'-t'') B_\beta E_\gamma(t') E_\delta^*(t'') + \text{c.c.} \qquad (1)$$

Here, $\Phi_{\alpha\beta\gamma\delta}$ is the fourth rank tensor whose non-vanishing components are determined by the point-group symmetry, **B** is the magnetic field, $\mathbf{E}(t)$ and $\mathbf{E}^*(t)$ are the (complex) amplitude of the electric field of the optical beam and its complex conjugate, respectively. In order to keep the further analysis simple, we restrict ourselves to the point-group symmetry of GaAs and choose the coordinate frame $x = [1\bar{1}0]$, $y = [110]$, and $z = [001]$. The optical propagation direction is along z. Moreover, to avoid any Hall currents we are only interested in the MCs parallel to the magnetic field **B**. Symmetry analysis of Eq. (1) shows that, for optical beams of fixed polarization states defined by the unit vector $\mathbf{e} = \mathbf{E}(t)/|\mathbf{E}(t)|$, the polarization dependence of MCs generated in the *x-y* -plane is given by[13]

$$j_{\text{MC}_{x,y}}(t) = -[\beta_l(t) \pm \sigma_l(t)] B_{x,y} (e_x e_y^* + \text{c.c.}) + [\sigma_c(t) \pm \beta_c(t)] B_{x,y} (e_x e_y^* - \text{c.c.}). \quad (2)$$

While $\beta_l$ and $\beta_c$ describe the photocurrents for the $T_d$ point-group symmetry of the bulk crystal (BMCs), $\sigma_l$ and $\sigma_c$ express the surface-related photocurrents for the effective $C_{\infty v}$ symmetry (SMCs). While the intensity of the optical excitation is included in the tensors, their subscripts indicate linearly (*l*) or circularly (*c*) polarized excitation. In the following we will refer to SMC (BMC) generated by linear and circular polarizations as LSMC (LBMC) and CSMC (CBMC), respectively.

One can see from the expression $(e_x e_y^* + \text{c.c.})$ of Eq. 2 that the amplitudes of the LSMC and LBMC are maximized for polarizations along the principle axes [100] and [010], i.e., for an angle of 45 ° with respect to x and y. Moreover, if one switches from one polarization state to the other, a sign change occurs for both currents. A current reversal is also obtained for helicity changes in the case of the CSMC and CBMC, which are maximized for circularly polarized light. Surface and bulk symmetry contributions can now be extracted from measurements of MCs using Eq. 2. If one measures $j_{\text{MC}_x}$ and turns the sample by ninety degrees without changing anything else one expects no changes in the SMCs, whereas the BMC contributions will change their sign. Therefore adding (subtracting) $j_{\text{MC}_x}$ and $j_{\text{MC}_y}$ will eliminate bulk (surface) contributions to the MC.

Before addressing the experimental results we note three issues:

(i) The measurements are influenced by unwanted signal contributions from other effects, like magnetic-field-independent bulk and surface photogalvanic effects,[1,4] photon drag effect, and others.[19] While most of these effects are symmetry forbidden in the employed geometry, some can still contribute to the measured THz traces as a background signal or due to imperfections of the experimental setup. To eliminate these contributions we changed the excitation polarization to invert the MCs as described above and subtracted the measured THz traces from each other. The same difference technique was performed for measured THz traces with positive and negative magnetic field, unless mentioned otherwise.

(ii) In certain geometries it is also possible to measure single MC contributions, with other contributions being symmetry forbidden. One example for such geometry would be the measurement of MCs along the principle axis [100]. The drawback for such a setup is that the LBMC only vanishes for specific optical polarization states, but does not reverse its sign, making the above described difference technique less effective. Moreover, in case of CBMC measurements a magnetic field perpendicular to the currents is required. This Hall geometry produces a strong background current, complicating the signal extraction. Besides these arguments, measurements along the principal axes also require more complicated setups in which either the magnetic field or the polarization direction of the detection have to be rotated. This is not necessary for our approach, which we believe provides the most reliable results.

(iii) Under the influence of magnetic fields the polarization of an optical beam can be changed due to the Faraday effect and magnetic-field induced birefringence. While the Faraday effect is geometrically forbidden in our setup, the magnetically induced birefringence is only relevant for above-bandgap excitation with $2 \cdot 10^{-4}$ deg/µm/T in GaAs[20] and, thus, can be ignored for very short optical absorption lengths of a few micrometer or less.[21]

*Experimental results* - The first set of experiments was performed at a photon energy of 1.57 eV. Figure 2(a) shows the THz traces of the MCs along the x- (black) and y-direction (red) for linearly polarized excitation. It is clear that both traces are almost equal indicating that a rotation of the sample inflicts hardly any influence on the obtained signals. It is obvious that only surface symmetry contributions are responsible for these traces. In Fig. 2(b) the corresponding LSMC and LBMC have been extracted by summation and subtraction, respectively. The THz traces of the MCs along the *x*- and *y*-directions obtained for circularly polarized light show a significant different behavior, see Fig. 2(c). Among a sign reversal also amplitude variations are observed, indicating a dominant bulk symmetry contribution accompanied with a weaker surface symmetry contribution. The extracted CSMC and CBMC traces are shown in Fig. 2(d). These results demonstrate a successful implementation of our symmetry separation scheme.

The dependences of the MCs on the optically induced carrier density are plotted in Fig. 3(a). As plotting parameter we used the peak to peak signal of the THz traces as shown in the insets of Fig. 3. Because of the very weak signal we will ignore the LBMC in this section. In the double logarithmic plot the slopes of the CBMC and CSMC are constant resembling a power dependence of 0.86 and 0.66, respectively. In contrast the carrier density dependence of the LSMC shows additional saturation effects. Most likely, the sub-linear behavior of all MCs is caused to a certain extend by changes of the THz propagation though the optically generated carrier plasma in the GaAs.[22] However, this effect has to be identical for all MCs. Therefore, the different dependence of the MC amplitudes on the carrier density directly proves different microscopic origins of the MCs. For visualization of the dependence of MCs on the magnetic field we do not use the difference technique for oppositely oriented magnetic fields, such that a current reversal is obtained for reversed magnetic fields, see Fig. 3(b). In the range of ±1.2 T only a linear dependence is observed, as indicated by the solid lines. This proves that the amplitude dependence of the MCs on magnetic field is well described by Eq. (2).

As final experimental result we plot the MC dependence on optical excitation energy, see Fig 4. Each intensity graph shows one type of MC for photon energies ranging from 1.39 eV to 1.58 eV. In all graphs the x-axis indicates the photon energy, while the y-axis indicates the time of the THz traces. Note that the maximum THz amplitude is varying from to graph to graph. Throughout all graphs the contour lines are slightly bended upwards for higher photon energies. We believe that this is due to different dispersions of the pump and the probe arms (in particular due to the electro-optic detection crystal) and will be disregarded in the further discussion. A strong signal from LSMC is present at 1.57 eV, see Fig. 4(a). The signal decreases for decreasing photon energies down to 1.48 eV. Another maximum occurs for the LSMC at 1.45 eV. At the band edge the LSMC changes its sign and is still present below the band gap, before vanishing at about 1.40eV. The spectral dependence of the CSMC is quite different. It has a pronounced peak right at the band edge and decreases monotonically with increasing photon energy. At 1.47 eV a phase jump appears as indicated by a bending of the contour lines. Increasing the photon energy further the CSMC keeps its monotonic trend, crosses zero at 1.53 eV and rises again. The LBMC signals are very close to the noise level throughout the photon energy range, see Fig. 4(c), so we are not commenting on any further details. The CBMC plotted in Fig. 4(d) shows a quite similar behavior as compared to the LSMC, except for the energies at the band edge, where the phase shift of the LSMC appears to be different than π (which one would obtain for a current reversal).

*Discussions* - Comparisons of our results to previous experiments on MC are quite difficult since previous measurements were done under different experimental conditions, mainly in n-doped GaAs using time integrated measurements.[7,9] In the following we will still try to draw some similarities and point out differences. For this discussion it is important to keep in mind that the THz radiation is proportional to the time derivative of the currents. Thus we are mainly detecting fast current changes (i.e., the onset of the currents and the decay due to fast relaxation mechanisms) but no slow changes (i.e., due to spin relaxation or carrier recombination).

The LSMC is based on an optically induced anisotropic carrier distribution (optical alignment of momenta), which gets rotated in the magnetic field. The current itself is then created by asymmetric carrier scattering at the surface.[7] Carrier/LO-phonon scattering will reduce the optically induced anisotropy drastically.[6] Alperovich et al.[7] observed an oscillating behavior of the LSMC versus photon energy at liquid helium temperatures with time integrated measurement techniques, revealing influence of LO-phonon emission. The broadening of states at room temperature and the allowance to absorb phonons make the observation of oscillations for us very difficult. However, we still expect to see small contributions of LO-phonon emission to our signal and it is very likely that this causes the non-monotonic current dependence on photon energy (decrease at 1.46 eV, increase at 1.51 eV) suggesting that the LSMC originates from the conduction band.

The only known microscopic origin for the CSMC, we are aware of, is the inverse spin Hall effect.[9,12,15,16] This electric current is orthogonal to a flow of spin polarized carriers (induced by a spin gradient or built-in electric field at the surface) and a spin orientation.[9] Circularly polarized light arranges the spin of the optically excited carriers parallel or antiparallel to the optical wave vector. Due to the absorption a spatial spin gradient and, consequently, a flow of spin polarized electrons is created. An applied magnetic field rotates the carrier spin creating the desired angle between spin and spin flow, and thus, producing the CSMC.[12] The required angle might also be created by the Hall effect of the static magnetic field on the spin flow. Bakun et al.[9] measured a maximum of the CSMCs at 10 mT at liquid nitrogen temperatures, which was explained with the Hanle effect causing a faster spin relaxation at higher magnetic fields. However, the spin relaxation time is in the nanosecond regime, which is too slow to be detected in our measurement. This might be the reason why we just observe a linear dependence of the CSMC on magnetic field. In any case the current enhancement at the bandgap and the phase change in the continuum reveal complex carrier dynamics of the CSMCs.

Andrianov et al.[8] studied BMCs using intraband excitations with a $CO_2$ laser on p-doped GaAs at room temperature. The measured strength of the LBMC at 0.7 T was about 8 times weaker than the CBMC strength. Although we used very different photon energies and measurement techniques, the same qualitative result are obtained. For a photon energy of 1.58 eV and a magnetic field of 1.2 T the LBMC was not clearly separable from the noise, but it was at least 20 times weaker than the CBMC. Our observed change of the CBMC at the energy where LO-phonon emission in the conduction band is allowed (see also discussion for LSMC) suggests that the CBMC is originating from the conduction band. This is in line with calculations from Ivchenko et al.[14] showing that the CBMC can occur due to spin-orbit splitting of the conduction band ($k$-cubic terms in the Hamiltonian).

In summary, we showed that ultrafast optical interband excitation of bulk GaAs leads to MCs, which can be detected via the simultaneously emitted THz radiation at room temperature. Using symmetry considerations, surface and bulk contributions to the overall current response could be separated in a straightforward way. Our introduced measurement scheme is important for thorough comparison of MCs. Moreover, our results reveal complex dynamics of such currents, which cannot be explained with existing time-independent theories. Further investigations on MCs are in particular promising since they offer the possibility to study very different relaxation processes such as polar versus anisotropic momentum relaxation under the same experimental conditions.

The authors thank the "Deutsche Forschungsgemeinschaft" (DFG) for financial support.

# References

[1] V.I. Belinicher and B.I. Sturman, "The Photogalvanic Effect in Media Lacking a Center of Symmetry," Sov. Phys. Uspekhi **23**, 199 (1980).

[2] S.D. Ganichev and W. Prettl, "Spin Photocurrents in Quantum Wells," J. Phys. Condens. Matter **15**, R935 (2003).

[3] S.D. Ganichev, E.L. Ivchenko, V. V. Bel'kov, S. a. Tarasenko, M. Sollinger, D. Weiss, W. Wegscheider, and W. Prettl, "Spin-Galvanic Effect," Nature **417**, 2 (2002).

[4] J.E. Sipe and A.I. Shkrebtii, "Second-Order Optical Response in Semiconductors," Phys. Rev. B **61**, 5337 (2000).

[5] L.I. Magarill and M. V. Éntin, "Surface Photogalvanic Effect in Metals," Sov. Physics, JETP **54**, 531 (1981).

[6] V.L. Alperovich, V.I. Belinicher, V.N. Novikov, and A.S. Terekhov, "Photogalvanic Effects Investigation in Gallium Arsenide," Ferroelectrics **45**, 1 (1982).

[7] V.L. Alperovich, A.O. Minaev, and A.S. Terekhov, "Ballistic Electron Transport through Epitaxial GaAs Films in a Magnetically Induced Surface Photocurrent," JETP Lett. **49**, 702 (1989).

[8] A. V. Andrianov, E.V. V. Beregulin, Y.B. Lyanda-Geller, I.D.D. Yaroshetskii, and Y.B. Layanda-Geller, "Magnetic-Field-Induced Photovoltaic Effect in P-Type Gallium Arsenide," Sov. Physics, JETP **75**, 921 (1992).

[9] A.A. Bakun, B.P. Zakharchenya, A.A. Rogachev, M.N. Tkachuk, and V.G. Fleischer, "Observation of a Surface Photocurrent Caused by Optical Orientation of Electrons in a Semiconductor," JETP Lett. **40**, 1293 (1984).

[10] E.L. Ivchenko, Y.B. Lyanda-Geller, and G.E. Pikus, "Magneto-Photogalvanic Effects in Noncentrosymmetric Crystals," Ferroelectrics **83**, 19 (1988).

[11] S.B. Astaf'ev, V.M. Fridkin, V.G. Lazarev, and A.L. Shlensky, "Magnetophotovoltaic Effect in Crystals without a Center of Symmetry," Ferroelectrics **83**, 3 (1988).

[12] N.S. Averkiev and M.I. D'yakonov, "Current due to Inhomogeneity of the Spin Orientation of Electrons in a Semidurctor," Sov. Physics, Semicond. **17**, 393 (1983).

[13] V. V. Bel'kov, S.D. Ganichev, E.L. Ivchenko, S. a. Tarasenko, W. Weber, S. Giglberger, M. Olteanu, H.-P. Tranitz, S.N. Danilov, P. Schneider, W. Wegscheider, D. Weiss, and W. Prettl, "Magneto-Gyrotropic Photogalvanic Effects in Semiconductor Quantum Wells," J. Phys. Condens. Matter **17**, 3405 (2005).

[14] E.L. Ivchenko, Y.B. Lyanda-Geller, and G.E. Pikus, "Circular Magnetophotocurrent and Spin Splitting of Band States in Optically-Inactive Crystals," Solid State Commun. **69**, 663 (1989).

[15] K. Ando and E. Saitoh, "Observation of the Inverse Spin Hall Effect in Silicon.," Nat. Commun. **3**, 629 (2012).

[16] T. Jungwirth, J. Wunderlich, and K. Olejník, "Spin Hall Effect Devices," Nat. Mater. **11**, 382 (2012).

[17] P.C.M. Planken, H.-K. Nienhuys, H.J. Bakker, and T. Wenckebach, "Measurement and Calculation of the Orientation Dependence of Terahertz Pulse Detection in ZnTe," J. Opt. Soc. Am. B **18**, 313 (2001).

[18] D. Côté, N. Laman, and H.M. van Driel, "Rectification and Shift Currents in GaAs," Appl. Phys. Lett. **80**, 905 (2002).

[19] H. Füser and M. Bieler, "Terahertz Beam Steering by Optical Coherent Control," Appl. Phys. Lett. **102**, 251109 (2013).

[20] B.B. Krichevtsov, R. V Pisarev, A.A. Rzhevskii, and H.-J. Weber, "Manifestation of Magnetically Induced Spatial Dispersion in the Cubic Semiconductors ZnTe, CdTe, and GaAs," J. Exp. Theor. Phys. Lett. **69**, 551 (1999).

[21] J.S. Blakemore, "Semiconducting and Other Major Properties of Gallium Arsenide," J. Appl. Phys. **53**, R123 (1982).

[22] M. Bieler, K. Pierz, and U. Siegner, "Simultaneous Generation of Shift and Injection Currents in (110)-Grown GaAs/AlGaAs Quantum Wells," J. Appl. Phys. **100**, 083710 (2006).

Fig. 1: (a) Experimental setup with optical polarization control of the optical pump beam via λ/2 and λ/4 waveplates. A low pass filter (LPF) and THz-polarizer were placed in between the off-axis parabolic mirrors. (b) Sample with the measured current direction being parallel to the magnetic field. During the experiments the sample was rotated in the magnetic field as indicated by the curved arrow.

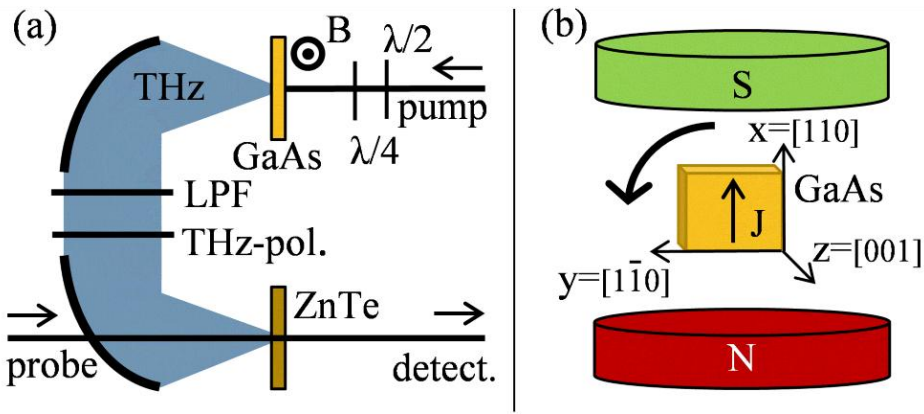

Fig. 2: Measured electrical THz traces from linearly (a) and circularly (c) polarized excitation with a photon energy of 1.57 eV. (b) and (d) show the extracted surface and bulk contributions. The unit of the electrical THz fields is equal to the intensity change of the probe beam obtained from electro-optic sampling.

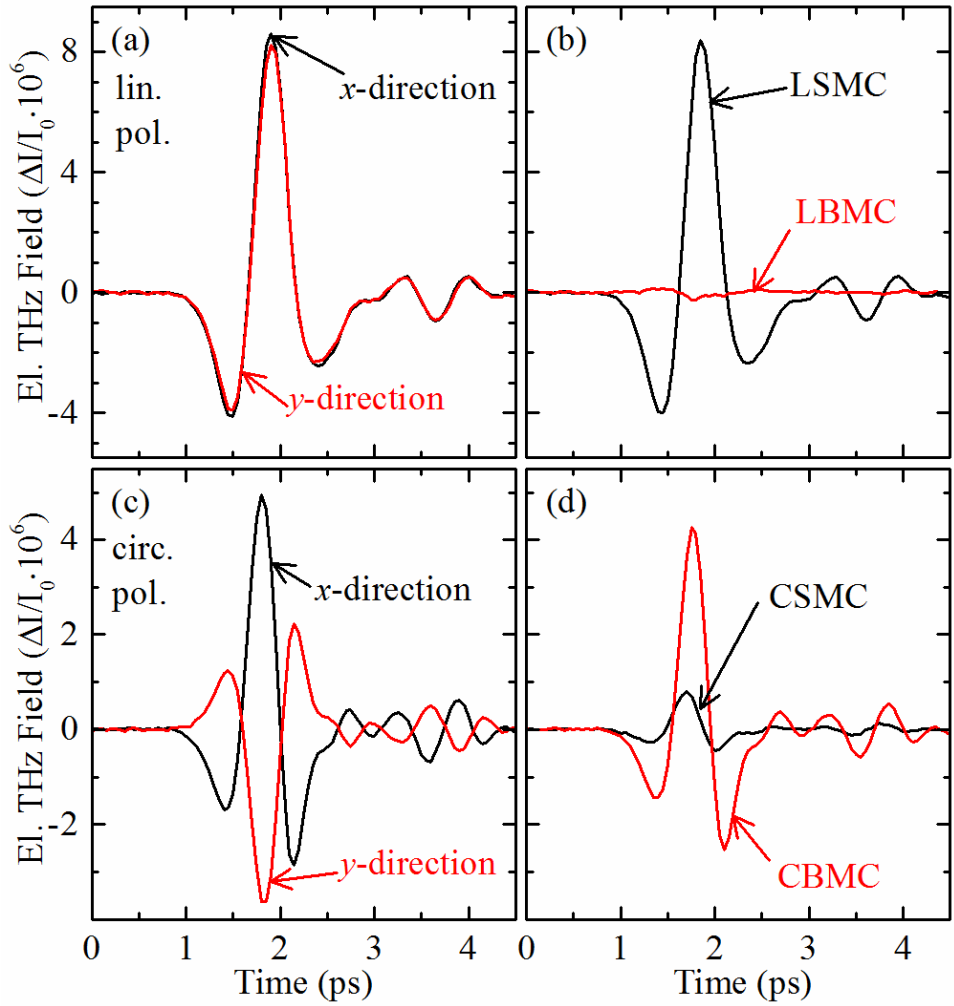

Fig. 3: Peak-to-peak (Pk-Pk) values of the THz traces emitted from MCs for 1.57 eV photon excitation energy versus (a) optically induced carrier density and (b) magnetic field. The solid lines in (b) are linear fits to the experimental data. The insets show the Pk-Pk extraction.

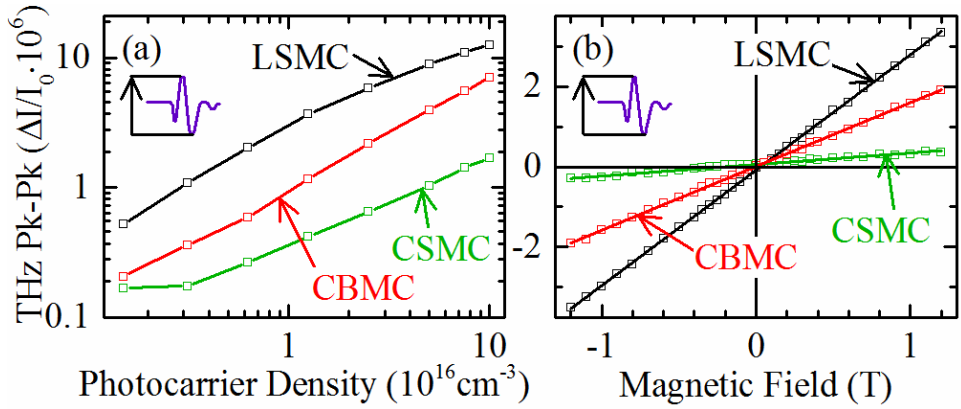

Fig. 4: Electrical THz traces emitted from MCs versus photon energy. The red pulses denote the spectral width of the optical excitation pulse. The black line indicates the band gap at room temperature.

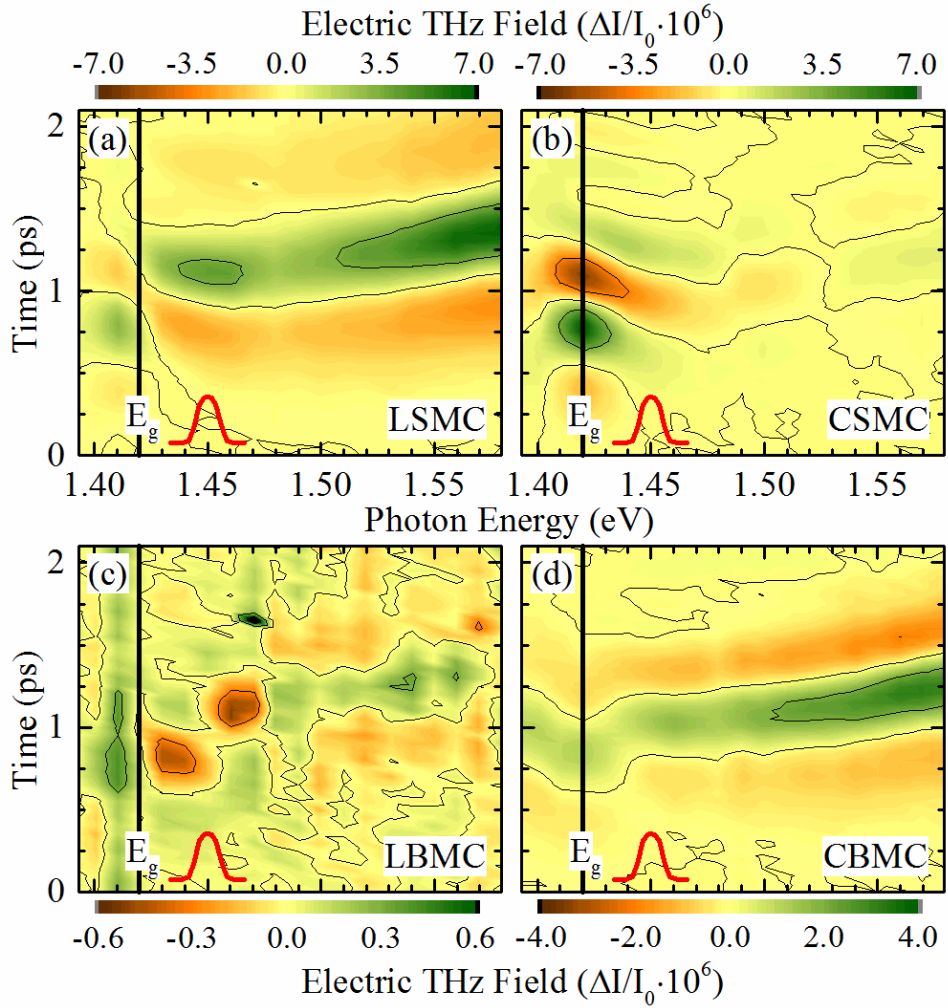